\documentclass[aps,prl,twocolumn]{revtex4-1}

\include{setspace}
\include{dcolumn}
\usepackage{graphicx}
\usepackage{longtable}
\usepackage{amsmath}
\usepackage{multirow}
\begin{document}

\title{Excitation of Giant Monopole Resonance in $^{208}$Pb and $^{116}$Sn Using Inelastic Deuteron Scattering}

\author{D. Patel$^{1,2}$,\, U. Garg$^{1,2}$,\, M. Itoh$^{3}$,\, H. Akimune$^{4}$,\, G.P.A. Berg$^{1,2}$,\, M. Fujiwara$^{5}$,\, M.N. Harakeh$^{6,7}$,\, \\ C. Iwamoto$^{4}$,\, T. Kawabata$^{8}$,\, K. Kawase$^{9}$,\, J.T. Matta$^{1,2}$,\, T. Murakami$^{8}$,\, A. Okamoto$^{4}$,\,T. Sako$^{9}$,\\ K.W. Schlax$^{1}$,\, F. Takahashi$^{5}$,\, M. White$^{1}$,\, and M. Yosoi$^{5}$\\}

\affiliation{$^1$Department of Physics, University of Notre Dame, Notre Dame, Indiana 46556, USA\\$^2$Joint Institute for Nuclear Astrophysics, University of Notre Dame, Notre Dame, Indiana 46556, USA\\ $^{3}$Cyclotron and Radioisotope Center, Tohoku University, Sendai 980-8578, Japan\\ $^4$Department of Physics, Konan University, Kobe 568-8501, Japan\\ $^{5}$Research Center for Nuclear Physics, Osaka University, Osaka 567-0047, Japan\\  $^{6}$Kernfysisch Versneller Instituut, University of Groningen, 9747 AA Groningen, The Netherlands\\$^7$GANIL, CEA/DSM-CNRS/IN2P3, 14076 Caen, France\\ $^{8}$Division of Physics and Astronomy, Kyoto University, Kyoto 606-8502, Japan\\ $^{9}$Japan Atomic Energy Agency, Kyoto 619-0215, Japan}

\date{\today}

\begin{abstract}
The excitation of the isoscalar giant monopole resonance (ISGMR) in $^{116}$Sn and $^{208}$Pb has been investigated using small-angle (including $0^\circ$) inelastic scattering of  100 MeV/u deuteron and multipole-decomposition analysis (MDA). The extracted strength distributions agree well with those from  inelastic scattering of 100 MeV/u $\alpha$ particles. These measurements establish deuteron inelastic scattering at E$_d \sim$ 100 MeV/u as a suitable probe for extraction of the ISGMR strength with MDA, making feasible the investigation of this resonance in radioactive isotopes in inverse kinematics. 
\end{abstract}

\maketitle

The nuclear incompressibility, K$_\infty$, is the measure of curvature of the equation of state (EOS) of nuclear matter at saturation density \cite{BohrMott1975}. The centroid energy of the isoscalar giant monopole resonance (ISGMR), a compression mode of nuclear vibrations, is directly related to the nuclear incompressibility of the finite nuclear matter, $K_A$ \cite{Strin82,trein81}, from which K$_\infty$ can be extracted \cite{blaizot, colo1,shlomo}, thereby providing an estimate for this important quantity. Further, from measurements of the ISGMR strengths in the series of stable isotopes of Sn ($Z$=50) and Cd ($Z$=48), the asymmetry term of the nuclear incompressibility, K$_\tau$, has been extracted \cite{Tao2007,Tao2010,DP2012}. This asymmetry term is directly related to the symmetry energy, E$_{sym}$, which is important in understanding the equation of state of dense asymmetric nuclear matter, as found in neutron stars, for example.

To further constrain the value of K$_{\tau}$ (and, consequently, E$_{sym}$), requires determination of ISGMR strengths in nuclei away from the valley of stability. Further interest in ISGMR strengths in nuclei far from stability, especially on the neutron-rich side, stems from the possible investigation of many important and intriguing nuclear structure effects, such as the theoretically
predicted appearance of monopole strength below the particle threshold: the ``pygmy'' ISGMR \cite{sagawa,khan,khan2}, akin to the ``pygmy'' dipole resonances reported in several neutron-rich nuclei (see, for example, Ref. \cite{aumann}). Because of the finite lifetimes of nuclei away from the valley of stability, such studies by necessity have to be performed in inverse kinematics where the radioactive nuclei are used to bombard target of lighter nuclei, e.g. helium or deuterium. The advent of new radioactive-ion beam facilities makes these investigations feasible. Inverse-kinematics measurements can then be made using an active target system (such as MAYA \cite{mittig} at GANIL, for example) wherein the detector gas also acts as the target. In the first such measurement, using MAYA, an enhancement of giant resonance (GR) strength was observed on top of the underlying continuum in the  $^{56}$Ni nucleus and it was demonstrated that this ``bump'' could be construed as corresponding to a combination of the ISGMR and the isoscalar giant quadrupole resonance (ISGQR)  \cite{Monrozeau2007,Monrozeau2008}. This measurement used deuterium as the active target/detector gas. As documented in Ref.\cite{Monrozeau2008}, it proved impossible to operate the counter with pure He gas, making it very difficult to use $\alpha$ particle scattering in inverse kinematics for ISGMR investigations in radioactive nuclei. To clearly establish the usefulness of deuterium as a probe of isoscalar giant resonances, it is important to validate in known cases the results obtained by inelastic deuteron scattering through comparison with reliable results obtained by inelastic alpha scattering.

As an isoscalar particle, the deuteron is ideally suited for the investigation of the ISGMR and some experimental work with this projectile was carried out in the 1970's \cite{marty, Willis}. Specifically, Willis {\em et al.} \cite{Willis} performed a (d,d$^\prime$) measurement at 54 MeV/u, where inelastic cross-sections for giant resonances are rather low. Moreover, the GR strength distributions for various multipoles were extracted using peak fitting, a method deemed less reliable than the multipole decomposition analysis (MDA) technique currently in use. The lack of prior knowledge of GR excitation using deuteron probe rendered the analysis difficult in the aforementioned $^{56}$Ni investigation as it was not possible to clearly delineate the ISGMR and ISGQR strengths. 

In this Letter, we present results of the first GR investigation with the deuteron probe at a beam energy amenable to reasonable cross sections for excitations of the GR. We have employed the now well-established MDA technique with ``instrumental-background-free'' inelastic scattering spectra \cite{bonin, Itoh2003}  and demonstrated, for the first time, that the ISGMR strength can be extracted reliably with this probe. This establishes inelastic deuteron scattering at forward angles as an attractive tool for future GR investigations in inverse kinematics with radioactive ion beams which are available at facilities currently in operation or being planned around the world.

The experiment was performed at the ring cyclotron facility of the Research Center for Nuclear Physics (RCNP), Osaka University, Japan. A 196-MeV $^2$H$^+$ beam was incident on 10 mg/cm$^2$-thick, enriched ($>$95\%) isotopic targets of $^{116}$Sn and $^{208}$Pb. At this beam energy, the angular distributions of differential cross sections at small angles are quite distinct for each multipole, making a reliable extraction of various strength distributions possible with the MDA technique. Also, the equivalent beam energy is (will be) available at the present (forthcoming) radioactive ion-beam facilities. 

Both elastic and inelastic scattering data were obtained in the measurements being reported in this Letter. Elastic scattering cross-sections were measured from 3.5$^{\circ}$ up to 32$^{\circ}$, in order to obtain the optical-model parameters (OMPs) required to calculate the expected cross sections for different multipoles. Inelastic scattering cross-section  measurements were performed at seven different angular settings, from 0.7$^\circ$ to 11$^\circ$. The measurement at the extremely forward angle of 0$^\circ$, corresponding to an average angle of 0.7$^\circ$ because of the finite angular bin for the measurement, is critical since the ISGMR cross-section is maximum at 0$^\circ$. In addition, the angular distributions for the various multipoles are most distinct at forward angles. 

\begin{figure}
\includegraphics[width=0.49\textwidth]{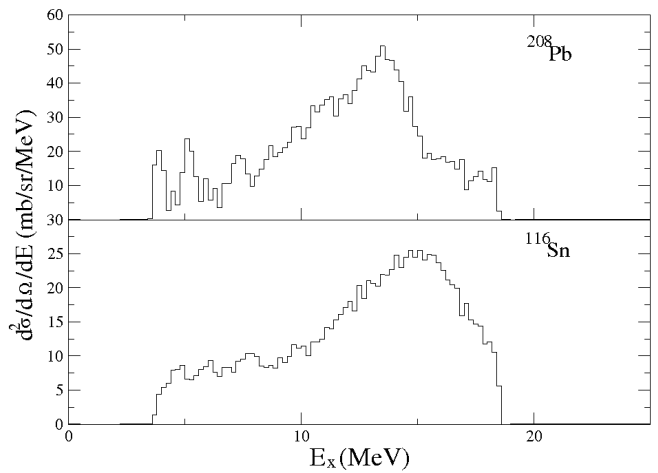}
\caption{Excitation-energy spectra from inelastic deuteron scattering for $^{116}$Sn and $^{208}$Pb at an incident energy of 196 MeV and an ``average'' scattering angle of 0.7$^\circ$. \label{fig:0-deg}}
\end{figure}

The scattered particles were momentum analyzed by the magnetic spectrometer Grand Raiden, and focused onto the focal-plane detector system \cite{Fujiwara99} which consisted of two multi-wire drift chambers (MWDC) and two plastic scintillators \cite{Itoh2003} allowing for identification of the scattered particles by time-of-flight and energy-loss techniques. For the  0$^{\circ}$ measurement, a special provision was made to let the main beam, which is very close to the scattered beam, pass through the detector system; details of the setup can be found in Ref. \cite{itoh-rcnp}. The detector at the focal plane covered from 6.5 MeV to $\sim$19 MeV in excitation energy, limited by the momentum bite of the spectrometer. The particle track was reconstructed using the ray-tracing technique described in 
Refs. \cite{Tao2010, Itoh2003}. This allowed reconstruction of the scattering angle; the angular resolution of the
MWDCs, including the nominal broadening of scattering angle due to the emittance of the beam and the
multiple Coulomb-scattering effects, was about 0.15$^{\circ}$.  Energy spectra were obtained for each scattering angle by software subdividing the full angular opening into three parts, each corresponding to a solid angle of 0.42 msr. Grand Raiden was used in the double-focusing mode in order to identify and eliminate all instrumental background \cite{Itoh2003} and the ``background-free'' (d,d$^\prime$) cross-section spectra so obtained for $^{116}$Sn and $^{208}$Pb at 0.7$^{\circ}$ are presented in Fig.~\ref{fig:0-deg}. The energy resolution of the spectra was $\sim$120 keV, more than sufficient for investigation of giant resonance that have expected widths of several MeV.

\begin{table}[h!]
	\begin{center}
	\caption{Optical Model parameters for $^{116}$Sn and $^{208}$Pb obtained by fitting the elastic scattering data.}
		\begin{tabular}{|c|c|c|c|c|c|c|}
\hline
Target & V$_{vol}$ & r$_{0,v}$ & a$_{v}$ & W$_{vol}$ & r$_{0,wv}$ & a$_{wv}$\\
& MeV & fm & fm & MeV & fm & fm \\ 
\hline
$^{116}$Sn & 44.33 & 1.18 & 0.911 & 20.87 & 1.07 & 0.571\\

$^{208}$Pb & 48.54 & 1.18 & 0.938 & 20.59 & 1.20 & 0.361\\
\hline
 & W$_{surf}$ & r$_{0,ws}$ & a$_{ws}$ & V$_{ls}$ & r$_{0,ls}$ & a$_{ls}$ \\
& MeV & fm & fm & MeV & fm & fm\\ 
\hline
$^{116}$Sn & 7.0 & 1.12 & 1.09 & 2.11 & 0.93 & 1.11\\

$^{208}$Pb & 7.0 & 1.24 & 0.79 & 2.11 & 1.12 & 1.23\\
\hline

\end{tabular}
\end{center}
\end{table}

\begin{figure}[h!]
\includegraphics[width=0.49\textwidth]{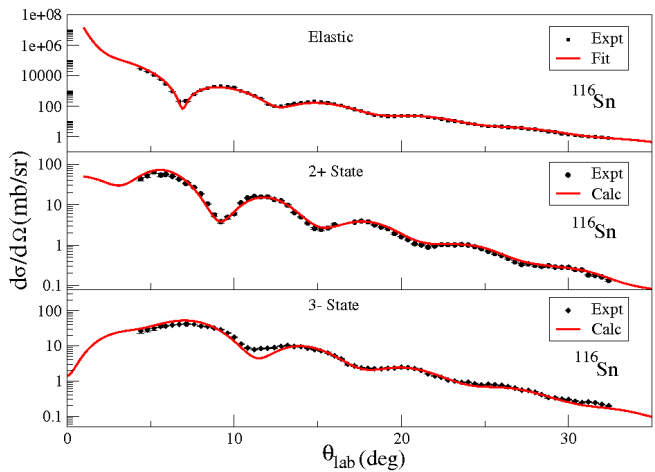}
\caption{(color online). Differential cross sections for deuteron elastic scattering off $^{116}$Sn at 196 MeV (top panel) and for excitations to the 2$^+_1$ (middle panel)  and 3$^-_1$ (bottom panel) states. The solid lines are results of DWBA calculations (see text). \label{fig:ElasticFit}}
\end{figure}

An optical-model potential was used in order to model the elastic scattering data \cite{Daehnick}. The potential form,
\begin{equation}
\label{eq:1}
V = V_{COUL}-V_{VOL}-iW_{VOL}+iW_{SURF}+V_{LS}
\end{equation}
was employed, where, 
\begin{equation*}
V_{VOL}=V_{vol}f_{v}(r,R_{v},a_{v}),
\end{equation*}
\begin{equation*}W_{VOL}=W_{vol}f_{wv}(r,R_{wv},a_{wv}),
\end{equation*}
\begin{equation*}
W_{SURF}=4a_{ws}W_{surf}{\frac{d}{dr}}f_{ws}(r,R_{ws},a_{ws}),
\end{equation*}
\begin{equation*}
V_{LS}=V_{ls}[\frac{\hbar}{m_{\pi}c}]^{2}[\vec{L}\cdot\vec{S}]\frac{1}{r}\frac{d}{dr}f_{ls}(r,R_{ls},a_{ls}),
\end{equation*}
and V$_{COUL}$ is the double-folded Coulomb potential obtained using the density-dependent double-folding calculation \cite{Khoa2000}. The functional form of ``{\em f}'' was chosen to be that of the Woods-Saxon potential. The computer code ECIS97 was employed to obtain the OMPs by fitting the elastic scattering angular distribution, using the $\chi^{2}$-minimization technique \cite{Raynal1988}; the parameters for $^{116}$Sn and $^{208}$Pb are listed in Table I. In order to validate the OMP set so obtained, experimental angular distributions of the differential cross-sections of the 2$^+_1$ and the 3$^-_1$ states in $^{116}$Sn and the 3$^-_1$ state in $^{208}$Pb were compared with the calculated DWBA differential cross-sections using adopted B(E2) and B(E3) values from Refs. \cite{nndc, Kibedi-spear}, as shown in Fig.~\ref{fig:ElasticFit} for the $^{116}$Sn case. A good agreement between the experimental and calculated angular distributions for the elastic, 2$^+_1$ and 3$^-_1$ states established the appropriateness of the OMPs. The OMPs were used in conjunction with the transition potentials and sum rules described in Refs. \cite{HarakehBook, Satchler87} to perform DWBA calculations for the various GRs.

\begin{figure}[h!]
\includegraphics[width=0.5\textwidth]{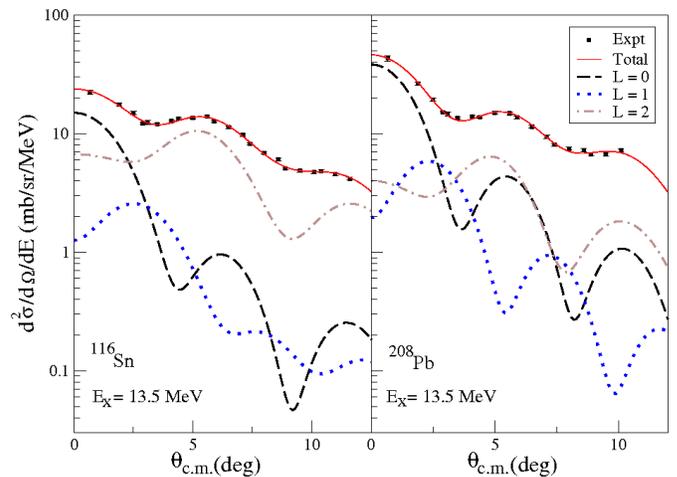}
\caption{(color online). MDA fits (solid red lines) for the differential cross sections obtained for the excitation-energy bins centered at 13.5 MeV in $^{116}$Sn (left panel) and $^{208}$Pb (right panel). While only the results for the L=0--2 contributions are displayed, the fits were performed for L=0--5 (see text). \label{fig:MDA}}
\end{figure}

The inelastic scattering spectra were divided into 1 MeV bins for further analysis. The experimental angular distribution of the differential cross-section at each excitation energy was fitted with a linear combination of the calculated DWBA differential cross-sections for various multipoles as follows:
\begin{equation}
\frac{d^2 \sigma^{exp}(\theta_{CM},E_x)}{d\Omega dE}=\sum_{L=0}^{L=5} a_L(E_x)\frac{d^2\sigma_L^{DWBA}(\theta_{CM},E_x)}{d\Omega dE}
\end{equation}
where, $L$ is the order of isoscalar multipole. The contribution of the isovector giant dipole resonance (IVGDR) to the total cross-section was subtracted out using photonuclear data in conjunction with the DWBA calculations based on the Goldhaber-Teller model \cite{Data}. Further details of the data analysis procedures will be provided in Ref. \cite{DP-thesis}. Typical MDA fits at 13.5 MeV in $^{116}$Sn and $^{208}$Pb, are shown in Fig.~\ref{fig:MDA}; the dominant contribution of the $L$=0 component can be seen clearly in the case of $^{208}$Pb; the $L$=2 component dominates, instead, for $^{116}$Sn.

\begin{figure}
\includegraphics[width=0.49\textwidth]{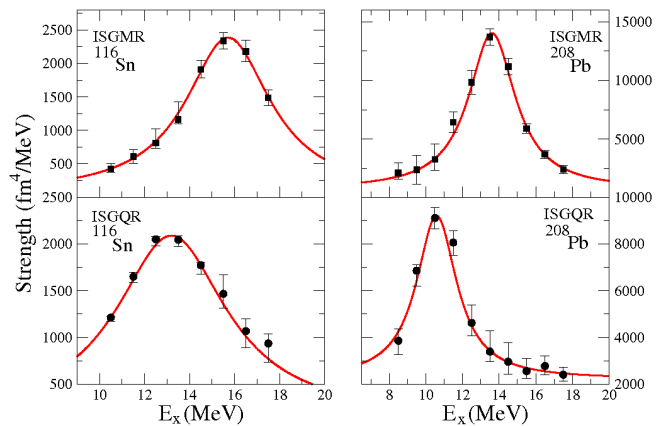}
\caption{(color online). ISGMR and ISGQR strength distributions for $^{116}$Sn and $^{208}$Pb obtained in this work. The solid red lines are Lorentzian fits to the data. 
\label{fig:strength}}
\end{figure}

\begin{table*}[!htb]
\caption{\label{hn} Lorentzian-fit parameters for the ISGMR and ISGQR strength distributions for $^{208}$Pb.  The ISGMR and ISGQR parameters from selected previous ($\alpha$,$\alpha^\prime$) measurements are also presented for comparison; the measurement labeled ``Orsay'' is from ($d$,$d^\prime$). Parameter values marked with an asterisk ($^\ast$) were extracted from the moment ratios rather than peak fitting. The EWSR values in this work are obtained over the full excitation energy range covered in this experiment; the quoted uncertainties are statistical only.}
\begin{tabular}{|c|c|c|c|c|c|c|c|c|}
\hline
 & \multicolumn{8}{|c|}{$^{208}$Pb} \\
\cline{2-9}
  &  & This work & Orsay \cite{Willis} & J\"{u}lich \cite{Morsch1980} & RCNP  \cite{Uchida2003}& TAMU$^\ast$ \cite{TAMU2004} & IUCF \cite{Bertrand1980} & KVI \cite{Harakeh1979}  \\
\cline{1-9}
ISGMR & E (MeV) & 13.6 $\pm$ 0.1 & 13.5 $\pm$ 0.3 & 13.8 $\pm$ 0.3 & 13.5 $\pm$ 0.2 & 13.96 $\pm$ 0.20 & 13.9 $\pm$ 0.4 & 13.9 $\pm$ 0.3 \\
\cline{2-9}
 & $\Gamma$ (MeV) & 3.1 $\pm$ 0.4 & 2.8 $\pm$ 0.2 & 2.6 $\pm$ 0.3 & 4.2 $\pm$ 0.3 & 2.88 $\pm$ 0.20 & 3.2 $\pm$ 0.4 & 2.5 $\pm$ 0.4 \\
\cline{2-9}
 & $\%$ EWSR & 147 $\pm$ 18 & 307 $\pm$ 60 & - & 58 $\pm$ 3 & 99 $\pm$ 15 & 100 $\pm$ 20 & 110 $\pm$ 22 \\
\cline{1-9}
ISGQR & E (MeV) & 10.6 $\pm$ 0.2 & 10.5 $\pm$ 0.2 & 10.9 $\pm$ 0.3 & - & 10.89 $\pm$ 0.30 & 10.9 $\pm$ 0.3 & 10.9 $\pm$ 0.3 \\
\cline{2-9}
 & $\Gamma$ (MeV) & 2.7 $\pm$ 0.4 & 2.8 $\pm$ 0.2 & 2.6 $\pm$ 0.3 & - & 3.00 $\pm$ 0.30 & 2.4 $\pm$ 0.4 & 3.0 $\pm$ 0.3 \\
\cline{2-9}
 & $\%$ EWSR & 98 $\pm$ 9 & 85 $\pm$ 15 & - & - & 100 $\pm$ 13 & 77 $\pm$ 15 & 145 $\pm$ 30 \\
\cline{1-9}
 \end{tabular}
\end{table*}

\begin{table*}[!htb]
\caption{\label{hn} Same as Table II, but for $^{116}$Sn.
%Lorentzian-fit parameters for the ISGMR and ISGQR strength distributions for $^{116}$Sn.  The ISGMR and ISGQR parameters from selected previous measurements using $\alpha$ probe are also presented for comparison. Parameter values marked with an asterisk ($^\ast$) were extracted from the moment ratios rather than peak fitting. The EWSR values in this work are obtained over the full excitation energy range covered in this experiment; the quoted uncertainties are statistical only.
}
\begin{tabular}{|c|c|c|c|c|c|}
\hline
 & \multicolumn{5}{|c|}{$^{116}$Sn} \\
\cline{2-6}
  &  & This work  & RCNP \cite{Tao2010} & TAMU$^{\ast}$ \cite{TAMU2004} & KVI \cite{Sharma1988}  \\
\cline{1-6}
ISGMR & E (MeV) & 15.7 $\pm$ 0.1 & 15.8 $\pm$ 0.1 & 15.85 $\pm$ 0.20 & 15.69 $\pm$ 0.16 \\
\cline{2-6}
 & $\Gamma$ (MeV) & 4.6 $\pm$ 0.7 & 4.1 $\pm$ 0.3 & 5.27 $\pm$ 0.25 & 3.73 $\pm$ 0.39 \\
\cline{2-6}
 & $\%$ EWSR & 73 $\pm$ 15 & 99 $\pm$ 5 & 112 $\pm$ 15 & 101 $\pm$ 22 \\
\cline{1-6}
ISGQR & E (MeV) & 13.2 $\pm$ 0.1 & 13.1 $\pm$ 0.1 & 13.50 $\pm$ 0.35 & 13.39 $\pm$ 0.14 \\
\cline{2-6}
 & $\Gamma$ (MeV) & 6.0 $\pm$ 1.0 & 6.4 $\pm$ 0.4 & 5.00 $\pm$ 0.30 & 2.94 $\pm$ 0.31 \\
\cline{2-6}
 & $\%$ EWSR & 73 $\pm$ 23 & 112 $\pm$ 4 & 108 $\pm$ 12 & 134 $\pm$ 28 \\
\cline{1-6}
 \end{tabular}
\end{table*}

The ISGMR and ISGQR strength distributions obtained from MDA for $^{116}$Sn and $^{208}$Pb are presented in Fig.~\ref{fig:strength}. The strength distributions are fitted with a Lorentzian functional form. The parameters extracted from these fits, as well as the corresponding results from selected previous studies using $\alpha$ and deuteron probes \cite{Tao2007,Willis,Morsch1980,Uchida2003,TAMU2004,Bertrand1980,Harakeh1979,Sharma1988} are presented in Table II and Table III for $^{208}$Pb and $^{116}$Sn, respectively. Incidentally, the parameters of the other compressional-mode, the isoscalar giant dipole resonance (ISGDR), cannot be extracted in any meaningful way in the present measurements because only a very small part of the ISGDR strength distribution is covered within the experimental excitation energy range. The properties of ISGMR and ISGQR extracted in this work agree very well with the previous values, however. This clearly establishes that it is possible to reliably extract ISGMR strength distributions from inelastic deuteron scattering using the MDA technique. 

To summarize, we have measured ISGMR strength in $^{116}$Sn and $^{208}$Pb using small-angle (including 0$^{\circ}$) inelastic deuteron scattering at a beam energy of 100 MeV/u. For the first time, it has been demonstrated that ISGMR strengths can be reliably extracted from ($d,d^\prime$) reactions using the MDA technique. Small-angle deuteron inelastic scattering can thus serve for reliable investigation of the ISGMR in nuclei far from stability using inverse-kinematics reactions, making it possible to investigate the properties of ISGMR in the exotic nuclei at the rare isotope beam facilities currently operational and being planned worldwide. 

We acknowledge the efforts of the RCNP staff in providing high-quality deuteron beams required for these measurements. This work has been supported in part by the National Science Foundation (Grants No. PHY-1068192 and No. PHY-0822648).

%\newpage

\end{document}